\RequirePackage{silence}
\documentclass[
  10pt,
  aps,
  pra,  
  onecolumn,
  superscriptaddress,
  longbibliography
]{revtex4-2}

\makeatletter
\def\pre@bibdata{group2_2Notes}
\makeatother

\usepackage[T1]{fontenc}
\usepackage[english]{babel}

\usepackage{amsmath,amssymb,amsfonts}
\usepackage{amsthm}
\usepackage{physics}
\usepackage{bbold}
\usepackage{upgreek}

\usepackage{graphicx}
\usepackage{tikz}

\usepackage{xcolor}

\DeclareUnicodeCharacter{2215}{/}

\usepackage[
  colorlinks=true,
  linkcolor=blue,
  citecolor=blue,
  urlcolor=blue
]{hyperref}

\begin{document}

\title{Pseudo Entropy in Quantum Spin Chains: from Integrability to Chaos}

\author{Tara Bahadur Rana}
\affiliation{St. Xavier's College, Tribhuvan University, Maitighar, 44600 Kathmandu, Nepal}
\affiliation{Holographic Himalaya, Lambagar, 44600 Tarakeshwar, Nepal}

\author{Yadav Raj Dahal}
\affiliation{Central Department of Physics, Tribhuvan University, Kirtipur, 44618 Kathmandu, Nepal}
\affiliation{Holographic Himalaya, Lambagar, 44600 Tarakeshwar, Nepal}

\author{Kiran Adhikari}
\email{kiran.adhikari@tum.de}
\affiliation{Emmy Noether Group for Theoretical Quantum Systems Design,
  Technical University of Munich,
Arcisstraße 21, 80333 München, Germany}

\begin{abstract}
  In this work, we study pseudo entropy and spectral dynamics in quantum spin chains (Heisenberg XXZ model and the mixed-field Ising model) to investigate whether they can serve as diagnostics of quantum chaos. First, we derive an exact relation between the real part of the thermal pseudo entropy and the spectral form factor. This provided evidence that pseudo entropy could exhibit the characteristic dip, logarithmic ramp, and plateau, implying the precise manner in which pseudo entropy probes chaos. We found that the imaginary part, while it doesn't diagnose chaos, can still provide crucial information about Fisher zeros that isn't available in the real counterpart. For spatial pseudo entropy in the XXZ chain, we find that the logarithmic critical scaling persists in both integrable and chaotic regimes, and that the non-positivity conjecture holds. Finally, we derive an exact connection between pseudo entropy, relative pseudo entropy, and spectral complexity with a potential connection to holography.
\end{abstract}

\maketitle
\section{Introduction}\label{intro_chaos}

Chaos has emerged as the third type of motion, involving deterministic dynamical systems that exhibit high sensitivity to initial conditions \cite{pub.1049937433,
goldstein2002classical, article, gershenson2004introductionchaosdeterministicsystems, PARK2026102133,Adhikari:2025zoa}. With the advent of quantum mechanics, one of the important problems in the field of quantum chaos is the question of how a quantum system behaves when its classical counterpart is chaotic \cite{1992Natur.355..311J}. A close relation between quantum chaos and random matrix theory (RMT) has been found and is widely accepted today \cite{PhysRevLett.52.1}. In RMT, chaotic systems follow the Wigner-Dyson level spacing distribution, while integrable systems follow Poisson statistics \cite{wigner1957statistical, PhysRevB.64.245336, Craps:2026vbm}. In the field of quantum chaos, RMT is used as a diagnostic tool for detecting chaos. Other diagnostic tools, such as the spectral form factor (SFF), spread complexity, K-complexity, and out-of-time-order correlators, have also been developed to
detect quantum chaos \cite{Rabinovici:2022beu,Hashimoto:2023swv, PhysRevE.109.054209,
10.21468/SciPostPhys.2.1.003, PhysRevResearch.7.023028, Garcia-Mata:2022voo}.

Recently, pseudo entropy has emerged as a novel tool in this diagnostic landscape \cite{He_2025}. Introduced in \cite{Nakata_2021} as a generalization of entanglement entropy, pseudo entropy is defined as the von Neumann entropy of a transition matrix, which is generally non-Hermitian, constructed between two distinct quantum states. Pseudo entropy was originally motivated by holography, which extends the geometric interpretation of entanglement entropy to Euclidean time-dependent spacetimes, where the usual notion of a minimal surface is not well defined \cite{Nakata_2021}. Subsequent studies have interpreted both the time-like entanglement entropy in AdS$/$CFT and the complex-valued holographic entanglement entropy in dS$/$CFT as pseudo entropy, with the imaginary part proposed to encode the emergence of a time direction in holography \cite{Doi:2022iyj, Doi:2023zaf}.

Beyond holography, pseudo entropy has found broad applications in many-body quantum systems \cite{Mukherjee:2022jac, He:2023eap, Mollabashi:2020yie, Mollabashi:2021xsd} and has been extended to topological phases through the concept of topological pseudo entropy \cite{Nishioka:2021cxe, Caputa:2024qkk}. It has also been explored in cosmology, particularly for two-mode squeezed states of cosmological perturbations \cite{Limbu:2026bol}. More recently, timelike entanglement measures have been shown to characterize ergodicity, scrambling, and spectral chaos effectively in random-matrix and many-body dynamical systems, with their imaginary parts interpreted as sensitive diagnostic signals \cite{Das:2026ifj, Das:2025fcd}. Pseudo entropy has likewise been investigated as a diagnostic of quantum chaos in the Sachdev--Ye--Kitaev (SYK) model \cite{He_2025}. Connections between pseudo entropy and the spectral form factor \cite{PhysRevD.104.L121902} suggest that pseudo entropy may encode signatures of chaotic dynamics.

Motivated by these developments, we study pseudo entropy in the quantum spin chains: Heisenberg XXZ and mixed-field Ising models, and investigate whether this quantity can detect the presence of quantum chaos. We identify a close relationship between the spectral form factor and pseudo entropy, both of which exhibit the characteristic slope-ramp-plateau structure. The linear ramp is present in chaotic systems but absent in integrable ones.

The paper is organized as follows. In section~\ref{PE}, we review the definition and properties of pseudo entropy, and in subsection~\ref{sec: SFF and pseudo entropy}, we develop a relation between the spectral form factor (SFF) and pseudo entropy. We introduce the quantum spin chains: Heisenberg XXZ and mixed-field Ising models in section~\ref{PE in spin chains}. In section~\ref{imaginary-part}, we shed light on the imaginary part of pseudo entropy via the use of thermal pseudo entropy. In section~\ref{sec:pseudo_entropy_complexity}, we introduce the relation between pseudo entropy and spectral complexity, and extend the relation to relative pseudo entropy and spectral complexity in subsection~\ref{sub sec: SC and RPE}. Finally, in section~\ref{sec:conclusion}, we summarize our results and outline future directions.

\section{Pseudo Entropy} \label{PE}

In quantum mechanics, the von Neumann entropy quantifies the degree of entanglement between quantum states (see \cite{RevModPhys.81.865} for a good review on quantum entanglement). It is defined by dividing the Hilbert space as $\mathcal{H} = \mathcal{H}_{A} \otimes \mathcal{H}_{B}$
and calculating

\begin{equation}
  S(\rho_{A}) = -\text{Tr}(\rho_{A}\text{log}\rho_{A}),
\end{equation}
where $\rho_{A} = \text{Tr}_{B}(\ket{\Psi}\bra{\Psi})$ is the reduced density matrix for the total quantum state $\ket{\Psi} \in \mathcal{H}$.

In the context of Holography, an analogous form of von Neumann entropy, called pseudo entropy, has been defined in \cite{Nakata_2021}. The pseudo entropy is a generalization of entanglement entropy calculated, à la von Neumann entropy,  by dividing the Hilbert space as $\mathcal{H} = \mathcal{H}_{A} \otimes \mathcal{H}_{B}$, and defining the reduced transition matrix of system $A$ by tracing out system $B$,
\begin{equation}
  \mathcal{T}^{\psi|\varphi}_A \equiv \mathrm{Tr}_B\left[\mathcal{T}^{\psi|\varphi}\right] = \mathrm{Tr}_B\left[\frac{|\psi\rangle\langle\varphi|}{\langle\varphi|\psi\rangle}\right].
  \label{eq:reduced}
\end{equation}
and pseudo entropy is defined as:
\begin{equation}
  \label{eq:Pseudo_entropy}
  S\!\left(\mathcal{T}^{\psi|\varphi}_A\right) = -\mathrm{Tr}\left(\mathcal{T}^{\psi|\varphi}_A \log \mathcal{T}^{\psi|\varphi}_A\right).
\end{equation}
For the case when the initial and final states are the same, i.e., $\ket{\psi} = \ket{\varphi}$, the pseudo entropy reduces to the ordinary entanglement entropy. While Eq. (\ref{eq:Pseudo_entropy}) looks like von Neumann entropy, the pseudo entropy is, in general, complex as $\mathcal{T}^{\psi|\varphi}_A$ is non-Hermitian with complex eigenvalues; only in some special choices of initial and final states is pseudo entropy real.

For calculation purposes, it is convenient to define the $n$-th R\'enyi entropy of the transition matrix $\mathcal{T}^{\psi|\varphi}_A$ just like we define the $n$-th R\'enyi entropy of a quantum state $\rho$:
\begin{equation}
  S^{(n)}\!\left(\mathcal{T}^{\psi|\varphi}_A\right) \equiv \frac{1}{1-n}\log\mathrm{Tr}\!\left[\left(\mathcal{T}^{\psi|\varphi}_A\right)^n\right],
  \label{eq:renyi}
\end{equation}
where $(n \in \mathbb{N}^+,\, n \geq 2)$, and we can simply choose the branch of the log function: $-\pi < \mathrm{Im}[\log(z)] \leq \pi$. We call this quantity $S^{(n)}(\mathcal{T}^{\psi|\varphi}_A)$ the pseudo $n$-th R\'enyi entropy, and taking the $n \to 1$ limit, one obtains the pseudo entropy:
\begin{equation}
  S\!\left(\mathcal{T}^{\psi|\varphi}_A\right) \equiv \lim_{n\to 1} S^{(n)}\!\left(\mathcal{T}^{\psi|\varphi}_A\right)
  = S\!\left(\mathcal{T}^{\psi|\varphi}_A\right).
  \label{eq:von}
\end{equation}

We can define a real-valued quantity, which has been relevant for phase transition studies \cite{Mollabashi:2021xsd, Kharel:2026puv}.
\begin{align}
  \Delta S^{(n)} \!\left(\mathcal{T}^{\psi|\varphi}_A\right) = \frac{1}{2} \Big[ &S^{(n)} \!\left(\mathcal{T}^{\psi|\varphi}_A\right) + S^{(n)} \!\left(\mathcal{T}^{\varphi|\psi}_A\right) -   S^{(n)}\!\left(\mathcal{T}^{\psi|\psi}_A\right) -   S^{(n)}\!\left(\mathcal{T}^{\varphi|\varphi}_A\right) \Big]
\end{align}
where, we note that $S^{(n)} \left(\mathcal{T}^{\psi|\varphi}_A\right) = S^{(n)} \left(\mathcal{T}^{\varphi|\psi}_A\right)^*$, and the latter two terms are the standard entanglement entropy for the state $\ket{\psi}$ and $\ket{\varphi}$ respectively. Taking the $n \to 1$ limit, we obtain $   \Delta S \left(\mathcal{T}^{\psi|\varphi}_A\right) =  \lim_{n\to 1} \Delta S^{(n)} \left(\mathcal{T}^{\psi|\varphi}_A\right).$
$ \Delta S \left(\mathcal{T}^{\psi|\varphi}_A\right)$ is thus nothing but the difference between the real part of the pseudo entropy and the averaged entanglement entropy:
\begin{align}
  \Delta S \!\left(\mathcal{T}^{\psi|\varphi}_A\right) = &\,\text{Re} \!\left(S\!\left(\mathcal{T}^{\psi|\varphi}_A\right) \right) - \frac{1}{2} \left( S(\rho_\psi)_A +  S(\rho_\varphi)_A \right)
  \label{eq: delta S}
\end{align}
where $ S(\rho_\psi)_A$ and $S(\rho_\varphi)_A$ is the standard entanglement entropy for the state $\ket{\psi}$ and $\ket{\varphi}$ of subsystem $A$. For the Lifshitz free scalar field and for Ising and XY spin models, it is seen that the difference $\Delta S \left(\mathcal{T}^{\psi|\varphi}_A\right)$ satisfies the following inequality:
\begin{equation}\label{inequality}
  \Delta S \left(\mathcal{T}^{\psi|\varphi}_A\right) \leq 0
\end{equation}
when the states $\ket{\psi}$ and $\ket{\varphi}$ are in the same phase, while if there is a quantum phase transition from  $\ket{\psi}$ to $\ket{\varphi}$, the inequality is typically violated, which is conjectured to be true for any QFT \cite{Mollabashi_2021}.

\subsection{Spectral form factor (SFF) and Pseudo entropy}\label{sec: SFF and pseudo entropy}

The Spectral form factor (SFF) is one of the central concepts in quantum chaos theory \cite{PhysRevE.55.4067, Cotler:2016fpe} defined via the analytically continued thermal partition function $Z(\beta + it) \equiv \Tr(e^{-(\beta + it)H})$ as
\begin{align}
  \label{eq:SFF}
  \mathrm{SFF}(\beta, t)
  &\equiv
  \frac{|Z(\beta+it)|^2}{Z(\beta)^2} =
  \frac{1}{Z(\beta)^2}
  \sum_{m,n}
  e^{-\beta(E_m+E_n)+i(E_m-E_n)t}
\end{align}
where, $\beta$ is the inverse temperature, and $\{E_n\}$ are the discrete eigenvalues of the Hamiltonian $H$. In quantum chaotic systems and random matrix theory, the spectral form factor $\mathrm{SFF}(t)$ exhibits a characteristic slope--dip--ramp--plateau structure, making it a powerful diagnostic of quantum chaos \cite{Das:2023yfj, PhysRevLett.134.010402, Camargo:2024deu, Cotler:2016fpe, PhysRevB.111.165108}. This behavior arises because random matrix spectra possess strong correlations between energy eigenvalues, leading to level repulsion and universal late-time spectral statistics. By contrast, integrable systems typically have uncorrelated energy levels and exhibit Poissonian level-spacing statistics. Since $\mathrm{SFF}(t)$ is constructed directly from the energy eigenvalues, it captures these spectral correlations and therefore distinguishes chaotic dynamics from integrable behavior.

To connect pseudo entropy with the SFF, we consider the thermofield double state (TFD), $\ket{\psi}$, as our initial state, and $\ket{\varphi}$ as the time-evolved version:
\begin{align}
  \ket{\psi}
  &= \frac{1}{\sqrt{Z(\beta)}}
  \sum_n e^{-\frac{\beta E_n}{2}}
  \ket{n}_L \otimes \ket{n}_R  \\
  \ket{\varphi}
  &= \frac{1}{\sqrt{Z(\beta)}}
  \sum_n e^{-\frac{\beta E_n}{2}} e^{iE_n t}
  \ket{n}_L\otimes \ket{n}_R.
\end{align}

The transition matrix is $\mathcal{T}^{\psi|\varphi} = \frac{\ket{\psi}\bra{\varphi}}{\braket{\varphi}{ \psi}},$ where the denominator is:
\begin{align}
  {\braket{\varphi}{ \psi}} &= \frac{1}{Z(\beta)} \sum_n e^{-\beta E_n} e^{-iE_n t}  \nonumber \\&= \frac{ \Tr(e^{-(\beta + it)H})}{Z(\beta)} \nonumber \\
  &= \frac{Z(\beta + it)}{Z(\beta)}
\end{align}
The reduced transition matrix of the right subsystem is:
\begin{align}
  \mathcal{T}_{R}^{\psi|\varphi} &= \text{Tr}_{L} \mathcal{T}^{\psi|\varphi} \nonumber \\
  &= \frac{1}{Z(\beta + it)} \sum_n e^{-(\beta + it)E_n} \nonumber \\
  &= \frac{e^{-(\beta+it)H_{R}}}{Z(\beta + it)}.
\end{align}
Hence, defining $z=\beta+it$, the pseudo entropy is the analytic
continuation of the thermal entropy,
\begin{align}
  S\!\left(\mathcal{T}^{\psi|\varphi}_R\right)
  &=
  z\langle H_R\rangle_z+\log Z(z)
  \nonumber\\
  &=
  \left(1-z\partial_z\right)\log Z(z),
  \label{eq:pseudo-complex-temp}
\end{align}
where
\begin{equation}
  \langle H_R\rangle_z
  =
  \frac{\Tr\!\left(H_R e^{-zH_R}\right)}{Z(z)}
  =
  -\partial_z\log Z(z).
\end{equation}
Now, we would like to take the real part of Eq.~\eqref{eq:pseudo-complex-temp}. For this, we can use the analyticity of $\log Z(z)$, and obtain an identity,
\begin{align}
  \mathrm{Re}\!\left[S\!\left(\mathcal{T}^{\psi|\varphi}_R\right)
  \right]
  &=
  \left(1-
    \beta\partial_\beta-t\partial_t
  \right)
  \log|Z(\beta+it)|   \nonumber \\
  &=
  \mathrm{Re}\!\left[
    (\beta+it)\langle H_R\rangle_{\beta+it}
  \right]
  +
  \log|Z(\beta+it)|
  \nonumber\\
  &=
  \mathrm{Re}\!\left[
    (\beta+it)\langle H_R\rangle_{\beta+it}
  \right]
  +
  \frac{1}{2}\log\mathrm{SFF}(\beta,t)
  +
  \log Z(\beta).
  \label{eq:ReS-decomp}
\end{align}
At infinite temperature, $Z(0)=d$, where $d$ is the Hilbert-space
dimension, and Eq.~\eqref{eq:ReS-decomp} reduces to
\begin{equation}
  \mathrm{Re}\,
  S\!\left(\mathcal{T}^{\psi|\varphi}_R\right)
  =
  \log d
  +
  \frac{1}{2}
  \left(
    1-t\partial_t
  \right)
  \log\mathrm{SFF}(0,t)
  .
  \label{eq:ReS-beta0}
\end{equation}
If the SFF has a power-law ramp
$\mathrm{SFF}(0,t)\propto t^a$, then
\begin{equation}
  \mathrm{Re}\,
  S\!\left(\mathcal{T}^{\psi|\varphi}_R\right)
  =
  \frac{a}{2}\log t+\mathrm{const},
  \label{eq:pseudo-ramp}
\end{equation}
so that the pseudo entropy exhibits a logarithmic ramp whose slope is fixed
by the SFF. A similar relation can be obtained for the pseudo-R\'enyi entropies.

\section{Pseudo entropy in Spin Chains}\label{PE in spin chains}

\subsection{The Heisenberg XXZ Model}

The spin-$1/2$ Heisenberg XXZ chain is a quintessential example for studying the transition from integrability to quantum chaos. The nearest-neighbor XXZ, $J_x=J_y\equiv J$ and $J_z\neq J$, model is integrable and can be solved by the Bethe ansatz
\cite{Bethe:1931hc,karbach1998introductionbetheansatzi,Karbach_1998,karbach2000introductionbetheansatziii}. Next-to-nearest-neighbor (NNN) interaction breaks Bethe-ansatz integrability, and drives the system to chaos:
\begin{align}
  H_{\rm XXZ}
  &=
  \sum_{i=1}^{N-1}
  \left[
    J\left(
      S_i^xS_{i+1}^x+
      S_i^yS_{i+1}^y
    \right)
    +
    J_zS_i^zS_{i+1}^z
  \right],
  \label{eq:xxz}
  \\
  H_{\rm NNN}
  &=
  \sum_{i=1}^{N-2}
  \left[
    J'\left(
      S_i^xS_{i+2}^x+
      S_i^yS_{i+2}^y
    \right)
    +
    J_z'S_i^zS_{i+2}^z
  \right], \\
  H&=H_{\rm XXZ}+\alpha H_{\rm NNN},
  \label{eq:nnn}
\end{align}
where $\alpha$ controls the strength of the NNN deformation \cite{Li_2008,10.1119/1.3671068,Vahedi_2016,Camargo:2024deu,Craps:2023rur} and,
\begin{equation}
  S_i^k
  =
  \left(\mathbb{1}_2\right)^{\otimes(i-1)}
  \otimes\sigma_k
  \otimes
  \left(\mathbb{1}_2\right)^{\otimes(N-i)},
  \qquad
  k\in\{x,y,z\},
\end{equation}
with $\sigma_k$ being the Pauli matrices. The Hilbert-space dimension is
$d=2^N$.
This model shows parity symmetry: $[H, \hat{\Pi}] = 0$,
\begin{equation}
  \hat{\Pi}=
  \begin{cases}
    \hat{P}_{1,N} \;\hat{P}_{2,N-1} \;...\;\hat{P}_{\frac{N}{2}, \frac{N+2}{2}}& \text{for $N$ even}\\
    \hat{P}_{1,N}\; \hat{P}_{2,N-1} \;...\;\hat{P}_{\frac{N-1}{2}, \frac{N+3}{2}}& \text{for $N$ odd}
  \end{cases}
\end{equation}
and
\begin{equation}
  \hat{P}_{i,j} = \frac{1}{2} (\mathbb{1}_{d} + S_{i}^{x}S_{j}^{x} + S_{i}^{y}S_{j}^{y} + S_{i}^{z}S_{j}^{z} ).
\end{equation}
The permutation operator permutes the spin configurations of the $i$th and $j$th sites. The action of $\hat{\Pi}$ on given spin configuration returns the mirror image; $\hat{\Pi} \ket{\uparrow \uparrow \uparrow \downarrow} = \ket{ \downarrow \uparrow \uparrow \uparrow}$. We can block-diagonalize the Hamiltonian into two sections of even and odd parity, labeled as $H_{1}$ and $H_{2}$, respectively. This symmetry resolution is essential when using level statistics as a
diagnostic of quantum chaos. Since both $M_z$ and parity are conserved,
nearest-neighbour level spacings must be evaluated within a fixed
$(M_z,\Pi)$ symmetry sector rather than from the full spectrum.

\begin{figure}[t]
  \centering
  \includegraphics[width=0.68\textwidth]{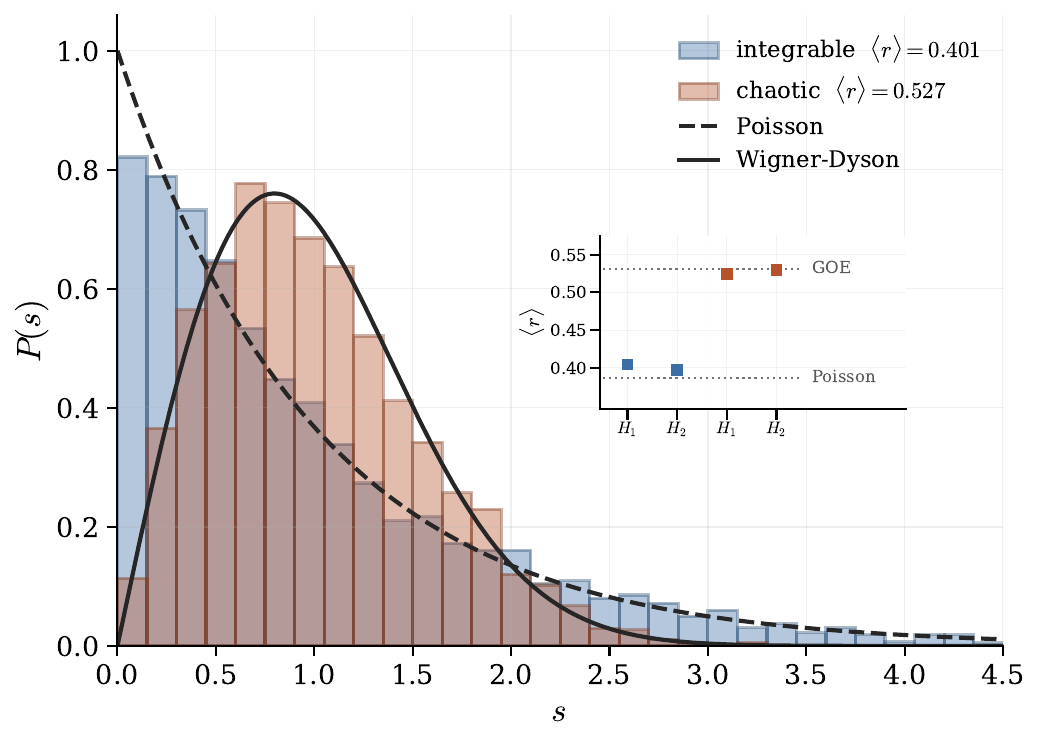}
  \caption{ Level spacing distribution $P(s)$ for the Heisenberg XXZ spin chain at $N=15$ in the magnetization sector $N_\uparrow=7$, with both
  reflection-parity sectors $H_1$ (even) and $H_2$ (odd) resolved and combined. The integrable chain, Eq.~\eqref{eq:xxz} with $J=1.0$, $J_z=0.5$, $\alpha=0$. The NNN-deformed chain, Eq.~\eqref{eq:nnn} with $J=J'=1.0$, $J_z=J_z'=0.5$, $\alpha=1$.}
  \label{fig:level-spacing}
\end{figure}

The resulting distributions are shown in Figure~\ref{fig:level-spacing}.
For $\alpha=0$, we obtain the Poisson level-spacing distribution, while tuning $\alpha=1$ gives the Wigner--Dyson form, the standard signature of quantum chaos \cite{MOSHFEGH2019502, PhysRevB.69.054403} with GOE statistics. In addition, we obtain the mean level-spacing ratio
$\langle r\rangle=0.404$ and $0.397$ in the even and odd sectors of the
integrable chain, and $\langle r\rangle=0.524$ and $0.530$ in the chaotic
chain, each consistent with the corresponding random-matrix prediction.

\subsubsection{Thermal Pseudo entropy}
\begin{figure}[t]
  \centering
  \includegraphics[width=\textwidth]{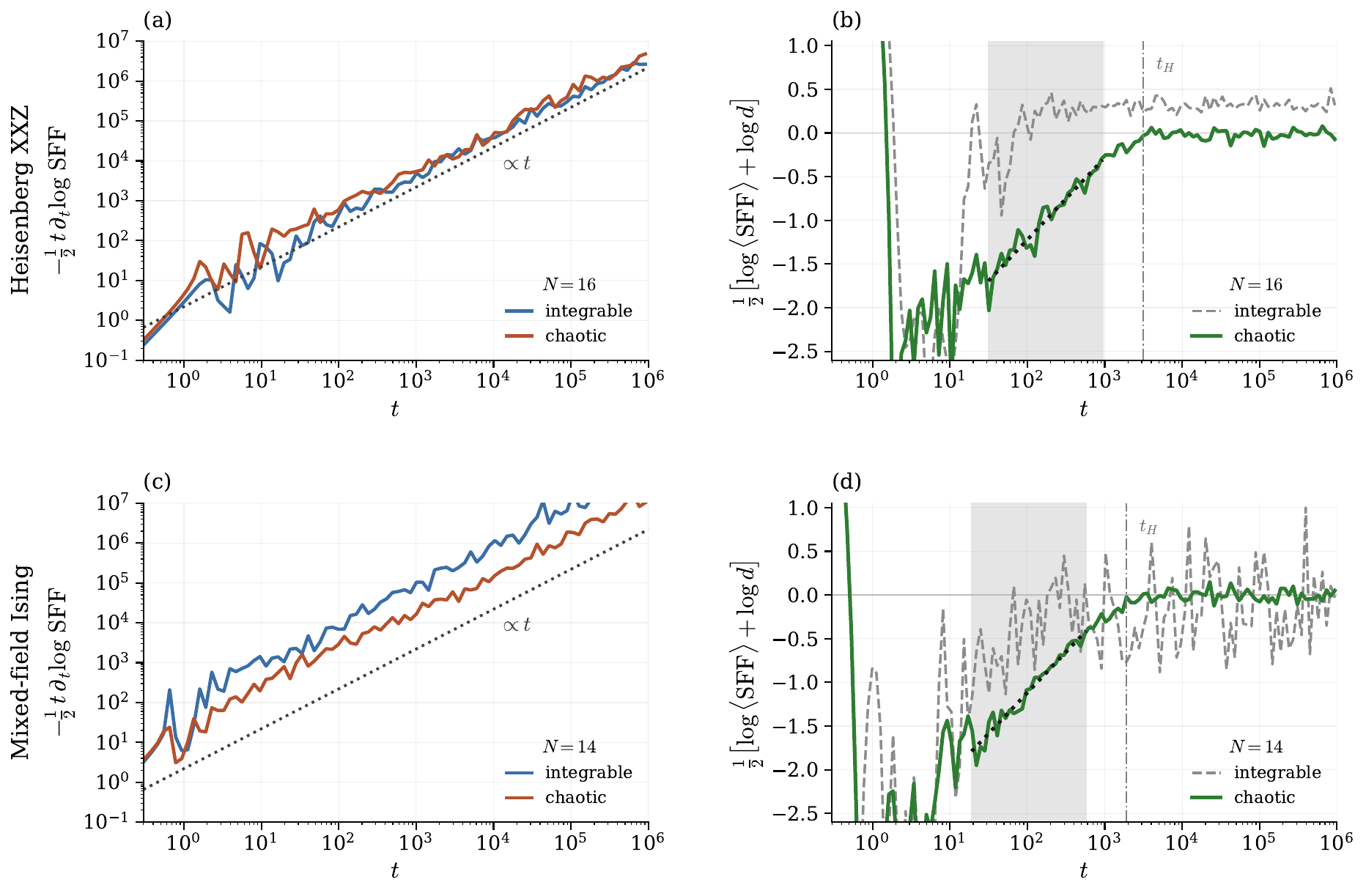}
  \caption{  Decomposition of the real pseudo entropy according to
    Eq.~\eqref{eq:ReS-beta0} at $\beta=0$ (infinite temperature), for the Heisenberg
    XXZ chain with $N=16$ (top row) and the mixed-field Ising chain with $N=14$
  (bottom row). Parameters: XXZ chain, $J=1.0$, $J_z=0.5$, with the chaotic case obtained by adding $\sum_i J_c S^z_i S^z_{i+2}$ at $J_c=1.0$; mixed-field Ising chain, $(h_x,h_z)=(-1.0,\,0.0)$ (integrable) and $(-1.05,\,0.5)$ (chaotic), both for open boundary conditions.}
  \label{fig:decomp}
\end{figure}

At temperature, $\beta = 0$, we saw that the Eq.~\eqref{eq:ReS-beta0} splits the real pseudo entropy into two pieces
\begin{equation}
  \underbrace{-\tfrac{1}{2}\,t\,\partial_t\log\mathrm{SFF}}_{\text{Term 1}}
  \;+\;
  \underbrace{\log d+\tfrac{1}{2}\log\mathrm{SFF}}_{\text{Term 2}}.
\end{equation}
The top row of Figure~\ref{fig:decomp}, Panel (a) for Term 1 and Panel (b) for Term 2, illustrates pseudo entropy for $N= 16$. Term 1 grows linearly in time for both the integrable and chaotic chains and provides no diagnostic information. A physical reason is that it is a one-point thermodynamic observable and is therefore insensitive to the two-point correlations between energy eigenvalues that are a hallmark of quantum chaos. A caveat is that we plot the envelope of Term 1 rather than its value, as it is dominated by large sign-alternating excursions due to zeros.

For the chaotic case, the Term 2 curve exhibits the characteristic dip, followed by an extended logarithmic ramp with a slope of $0.41$, followed by a plateau at the Heisenberg time $t_H$, consistent with the value $a/2 = 0.82/2$ predicted by Eq.~\eqref{eq:ReS-beta0}. In the integrable case, no such ramp exists: the form factor reaches its plateau at $t\approx20$ independently of the Hilbert space dimension $d$.

The form factor is averaged in logarithmic time bins before the
logarithm is taken. The chaotic curve shows the dip, ramp and plateau structure;
the shaded band marks the fit window $t\in[0.01,0.30]\,t_H$ and the dotted line
the corresponding least-squares fit, of slope $0.41$ in both models
($a=0.82$ and $a=0.81$ respectively). The dash-dotted vertical line marks the
Heisenberg time $t_H$, defined as the time at which the averaged form factor
first reaches its asymptotic value. The integrable curves display no ramp. Their
saturation value differs from the chaotic one for model-specific reasons: in
(b) the plateau lies $0.32$ higher because the exact degeneracies at
$\Delta=J_z/J=1/2$ raise the asymptotic form factor to $1.89/d$ rather than
$1/d$, whereas in (d) the integrable spectrum is non-degenerate and saturates at
$1/d$ but fluctuates strongly about it.

\subsubsection{Spatial pseudo entropy}
\label{sec:spatial-pseudo entropy}

So far, we have dealt with thermal pseudo entropy, constructed from the thermofield-double state. There is another possibility, spatial pseudo entropy,  constructed from two ground states $|\Psi_1\rangle,|\Psi_2\rangle$ of the XXZ chain and traced over a spatial subsystem $B$. This allows us to test the critical-scaling and non-positivity conjectures.

\begin{figure}[htbp]
  \centering
  \includegraphics[width=\textwidth]{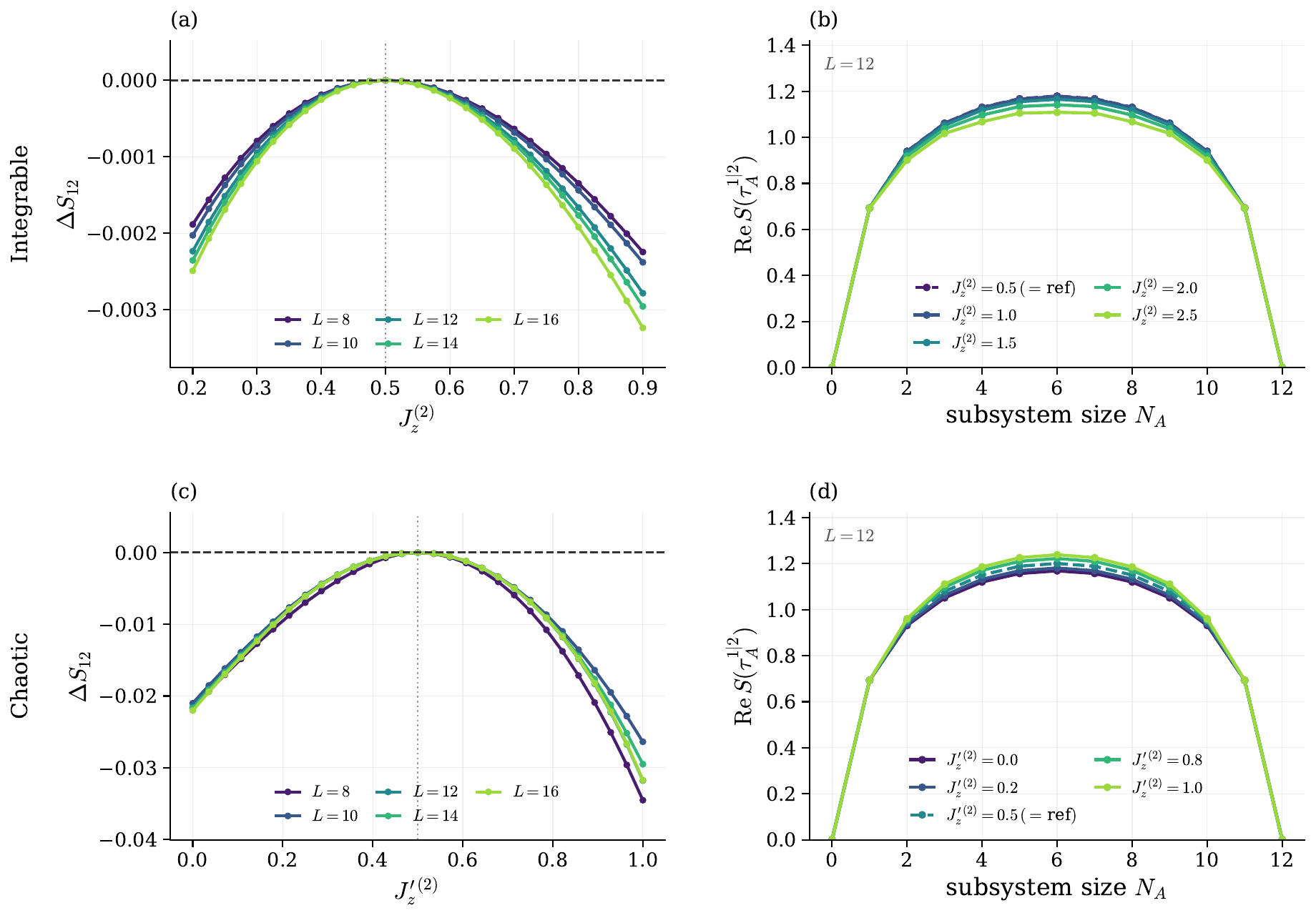}
  \caption{  Spatial pseudo entropy within the critical phase for $L=8$--$16$. (a) $\Delta S_{12}$ versus $J_z^{(2)}$, integrable case, $J_z^{(1)}=0.5$. (b) Pseudo entropy versus $N_A$, integrable case, $L=12$, for several $J_z^{(2)}$; the dashed curve is the self-comparison state. (c),(d) As in (a),(b) for the chaotic case, swept in $J_c^{(2)}$ at $J_z^{(1)}=J_z^{(2)}=0.5$.}
  \label{fig:spatial_nonpositivity}
\end{figure}

We take $|\Psi_1(J_z^{(1)})\rangle$ and $|\Psi_2(J_z^{(2)})\rangle$ for the integrable case Eq.~\eqref{eq:xxz}, and $|\Psi_1(J_z^{(1)},J_c^{(1)})\rangle$ and $|\Psi_2(J_z^{(2)},J_c^{(2)})\rangle$ for the chaotic case Eqs.~\eqref{eq:xxz} and~\eqref{eq:nnn}, under periodic boundary conditions in the $N_\uparrow=L/2$ sector. For $J_z^{(1)}=0.5$ and $L=12$, the pseudo entropy as a function of $N_A$ Figure~\ref{fig:spatial_nonpositivity} (b), (d) has a form
\begin{equation}
  S_A^{\rm pseudo}(N_A) \simeq c_1 \ln\!\left[\frac{L}{\pi}\sin\!\left(\frac{\pi N_A}{L}\right)\right] + c_0 .
  \label{eq:cft_fit}
\end{equation}
with $c_1\simeq0.311$--$0.360$, $c_0\simeq0.699$--$0.701$ in the integrable case and $c_1\simeq0.352$--$0.406$, $c_0\simeq0.699$--$0.700$ in the chaotic case. The almost identical values of $c_1$ indicate that the logarithmic critical scaling of entanglement entropy persists for pseudo entropy as well.

\begin{figure}[htbp]
  \centering
  \includegraphics[width=\textwidth]{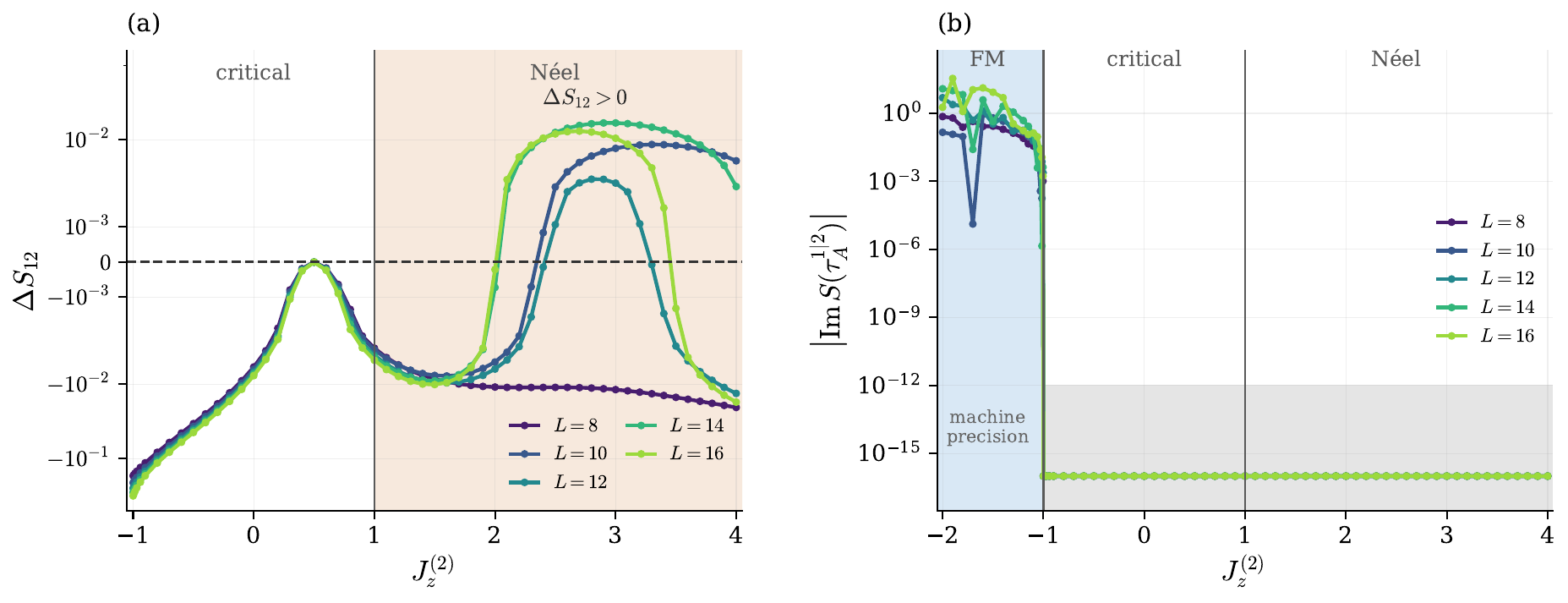}
  \caption {Spatial pseudo entropy across the XXZ phase diagram, integrable case, $L=8$--$16$, $J_z^{(1)}=0.5$. (a) $\Delta S_{12}$ versus $J_z^{(2)}$ across critical and N\'eel phases, on a symmetric-log scale (linear for $|\Delta S_{12}|<2\times10^{-3}$). The bound $\Delta S_{12}\le0$ holds throughout $-1<J_z^{(2)}\le1$ but is violated for $L\ge10$ at $J_z^{(2)}\gtrsim2.1$, with peak magnitude $\sim1.7\times10^{-2}$. (b) $|{\rm Im}\,S(\mathcal{T}_A)|$ versus $J_z^{(2)}$, showing the onset at the SU(2)-symmetric point $J_z^{(2)}=-1$.}
  \label{fig:spatial_transition}
\end{figure}

For the non-positivity test we evaluate $\Delta S_{12}$ Eq.~\eqref{eq: delta S} as a function of $J_z^{(2)}$ and $J_c^{(2)}$ at fixed $J_z^{(1)}=0.5$. Within the critical phase Figure~\ref{fig:spatial_nonpositivity} (a), (c), $\Delta S_{12}\le0$ for all $L=8$--$16$. The curve vanishes at the self-comparison point ($J_z^{(2)} = J_z^{(1)}=0.5$) where the pseudo entropy is identical to entanglement entropy. The curve then deepens monotonically on $L$, with the chaotic case roughly an order of magnitude deeper than the integrable one at fixed $L$. We then extend the analysis across the phases Figure~\ref{fig:spatial_transition} (a), and observe that $\Delta S_{12}$ becomes positive for $L\ge10$ once $J_z^{(2)}\gtrsim2.1$, so the bound fails. There is also an interesting observation. For $J_z^{(2)}<-1$, $S(\mathcal{T}_A)$ has a nonzero imaginary part Figure~\ref{fig:spatial_transition} (b) with a step-function like characteristics at $J_z^{(2)}=-1$. One mechanism behind this could be that $J_z^{(2)}=-1$ is an SU(2)-symmetric point where the Schmidt rank across the entangling cut collapses to its minimum.

\subsection{The mixed-field Ising Model}
We also introduce the mixed-field Ising model, which is a one-dimensional spin-$1/2$ chain. The mixed-field Ising model can be made integrable and chaotic simply by an appropriate choice of the longitudinal and transverse magnetic fields \cite{PhysRevLett.106.050405, PhysRevB.101.174313}.The Hamiltonian of the mixed-field Ising model includes both a longitudinal field $h_z$ and a transverse field $h_x$, and is written as \cite{Camargo:2024deu},
\begin{align}
  H &= -\sum_{i=1}^{N-1} S_i^z S_{i+1}^z - \sum_{i=1}^{N} \left( h_x S_i^x + h_z S_i^z \right),
  \label{equanising}\\
  S_i^k &= (\mathbb{1}_2)^{\otimes(i-1)} \otimes \sigma_k \otimes (\mathbb{1}_2)^{\otimes(N-i)}
  \label{equanspinising}
\end{align}

for $k \in \{x, y, z\}$, with $\sigma_k$ the usual Pauli matrices. The Hilbert space is $d = 2^N$-dimensional for $N$ sites. Like the XXZ spin chain, the model shows parity symmetry; that is, the Hamiltonian commutes with the parity operator
\begin{align}
  \hat{\Pi} &=
  \begin{cases}
    \hat{P}_{1,N}\, \hat{P}_{2,N-1} \cdots \hat{P}_{\frac{N}{2},\frac{N+2}{2}}, & \text{for } N \text{ even} \\[4pt]
    \hat{P}_{1,N}\, \hat{P}_{2,N-1} \cdots \hat{P}_{\frac{N-1}{2},\frac{N+3}{2}}, & \text{for } N \text{ odd}
  \end{cases}
  \label{eq:parity}\\
  \hat{P}_{i,j} &= \frac{1}{2}\left( \mathbb{1}_4 + S_i^x S_j^x + S_i^y S_j^y + S_i^z S_j^z \right)
  \label{eq:permutation}
\end{align}
The thermal pseudo entropy for the mixed-field Ising chain is plotted in Figure \ref{fig:decomp} (c and d). The overall behavior is identical to that of the Heisenberg XXZ Model, thus demonstrating another model in which pseudo entropy correctly diagnoses quantum chaos.

There are some technical points worth commenting on. Compared with the XXZ model, the mixed-field case exhibits distinct late-time behavior. In panel (b) of Figure \ref{fig:decomp} (XXZ model),  the integrable curve saturates approximately $0.32$ above the chaotic one. This is not a chaos signature but a
consequence of exact degeneracies: at $\Delta=J_z/J=1/2=\cos(\pi/3)$ the XXZ chain sits at a root of unity, where an enhanced symmetry beyond $U(1)\times\Pi$ leaves only $4222$ distinct levels among $6470$ (maximal degeneracy $6$) and raises the plateau of the form factor to $1.89/d$ rather than $1/d$. The same degeneracies displace $\langle r\rangle$ above the Poisson value in Figure~\ref{fig:level-spacing}. In panel (d)(Mixed field model), by contrast, the integrable
spectrum is entirely non-degenerate, and its plateau coincides with $1/d$, but the critical transverse-field Ising spectrum is sufficiently structured that the form factor fluctuates strongly about that value at all times.

\section{On the Imaginary Part: Thermal Pseudo Entropy}
\label{imaginary-part}

\begin{figure}[htbp]
  \centering
  \includegraphics[width=\textwidth]{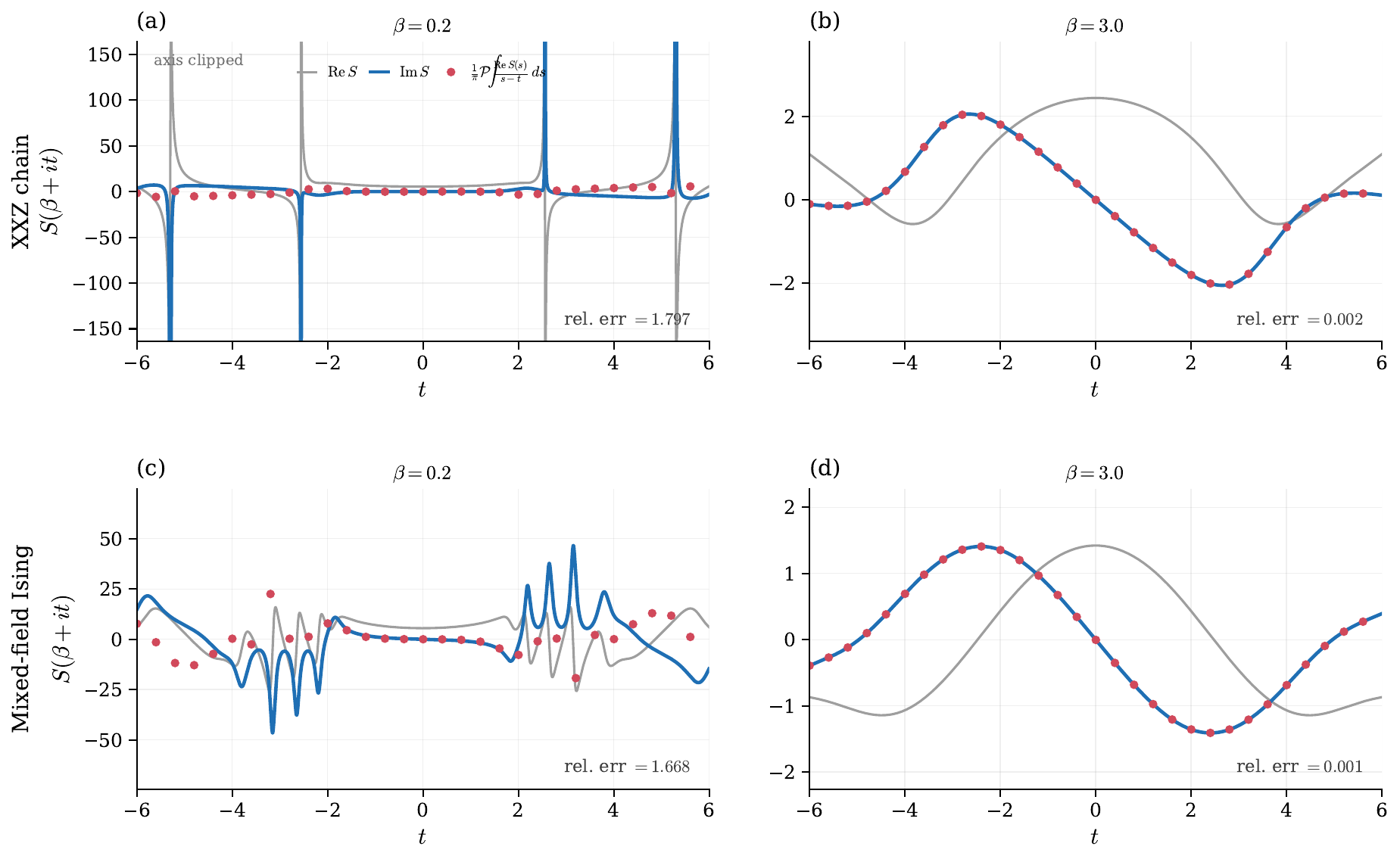}
  \caption{ The thermal pseudo entropy at $N=8$, evaluated on the continuously continued branch of $\log Z$. Red points: the Hilbert transform of $\Re S$, computed over $|s|\le400$. (a),(b) XXZ chain at $J_z=0.5$; (c),(d) mixed-field Ising
    chain. At $\beta=0.2$ Fisher zeros lie inside the dispersion contour,
  i.e.\ $\beta_k>\beta$ and the reconstruction fails. At $\beta=3.0$ the KK relation holds.}
  \label{fig:kk-relation}
\end{figure}

Although there have been significant developments in the study of the real part of the pseudo entropy, there hasn't been much progress in understanding the meaning of its imaginary part. In free scalar theory, the imaginary part of pseudo entropy is similar to the time derivative of the entanglement entropies \cite{Mollabashi:2021xsd}. Holographic pseudo entropy has been understood as the time-like entanglement entropy, with its imaginary part interpreted as the emergence of time in both AdS$/$CFT and dS$/$CFT \cite{Doi:2022iyj}. More recently, further dynamical interpretations of the
imaginary part have been proposed \cite{misumi2026imaginary,Misumi:2026jha,Xu:2024yvf}. Here we show that, for thermal pseudo entropy (TPE), the imaginary part contains nontrivial phase and winding information associated with Fisher zeros.

Recall that for a Hamiltonian $H$, where $z=\beta+it$, the thermal pseudo entropy Eq.~\eqref{eq:pseudo-complex-temp} is
\begin{equation}
  S(z)=   \left(1-z\partial_z\right)\log Z(z).
  \label{eq:tpe-exact}
\end{equation}
Even though $Z(z)$ is an entire function of $z$ (at least for a finite-dimensional system), the singularities of $S(z)$ can come from zeros and the branch structure of $\log Z(z)$. On any simply connected region containing no Fisher zero and no branch cut,
$S(z)$ is holomorphic, and Cauchy--Riemann equations imply
\begin{align}
  \frac{\partial^2u}{\partial\beta^2}
  +
  \frac{\partial^2u}{\partial t^2}
  &=0, \qquad
  \frac{\partial^2v}{\partial\beta^2}
  +
  \frac{\partial^2v}{\partial t^2}
  =0.
  \label{eq:laplace_v}
\end{align}
where, $S(\beta+it) = u(\beta,t)+i\,v(\beta,t)$. Thus, away from Fisher zeros, the real and imaginary parts of the TPE are locally harmonic conjugates. However, these are local statements and don't imply that the imaginary part is globally redundant. We therefore have to be careful near a Fisher zero.

Let $z_k$ be a Fisher zero of multiplicity $m_k$:
\begin{equation}
  Z(z)
  =
  (z-z_k)^{m_k}g_k(z),
  \qquad
  g_k(z_k)\neq0.
  \label{eq:fisher-factorization}
\end{equation}
Substitution into Eq.~\eqref{eq:tpe-exact} gives
\begin{align}
  S(z)
  &=
  m_k\log(z-z_k)
  -
  \frac{m_k z}{z-z_k}
  +
  \left(1-z\partial_z\right)\log g_k(z)
  \nonumber\\
  &=
  -\frac{m_k z_k}{z-z_k}
  +
  m_k\log(z-z_k)
  +
  S_k^{\mathrm{reg}}(z),
  \label{eq:tpe-near-zero}
\end{align}
where $S_k^{\mathrm{reg}}(z)$ is regular at $z=z_k$. The logarithmic term in Eq.~\eqref{eq:tpe-near-zero} introduces a nontrivial
monodromy around each Fisher zero. For a contour $\gamma_k$ encircling
$z_k$ once counterclockwise,
\begin{equation}
  \log(z-z_k)
  \longrightarrow
  \log(z-z_k)+2\pi i,
\end{equation}
whereas the pole and regular terms are single-valued. Consequently, the imaginary part carries a quantized winding around Fisher zeros, with
\begin{equation}
  m_k
  =
  \frac{1}{2\pi}
  \Delta_{\gamma_k}\operatorname{Im}S
  .
\end{equation}
Therefore, $\operatorname{Im}S$ can encode topological structures such as global winding numbers and multiplicity of Fisher zeros, not captured by the real counterpart. The connection between Fisher zeros, quantum criticality, and real-time
dynamics, and spectral statistics have been investigated in a number of
contexts
\cite{PhysRevResearch.6.043139,Bunin:2022kua,Lv:2026wza,
PhysRevLett.110.135704}. The above relations show that the imaginary TPE
provides a natural observable for studying this zero structure.

It was argued in \cite{Caputa:2024gve} that the standard Kramers--Kronig (KK) relation captures the relation between real and imaginary parts. However, it holds only if the relevant half-plane is free of Fisher singularities, and the required additional phase and winding information is provided by Fisher zeros. First, let us see when KK holds.

Let, $ F_\beta(w)=S(\beta+iw),$ with physical time corresponding to $w=t\in\mathbb{R}$. Since
$\beta+iw=(\beta-\Im w)+i\Re w$, the low-temperature limit
$\Re z\rightarrow+\infty$ corresponds to $\Im w\rightarrow-\infty$.
Thus, if $F_\beta(w)$ is analytic and sufficiently well behaved in the
lower half-plane, its boundary values satisfy the KK relation:
\begin{equation}
  \Im S(\beta+it)
  =
  \frac{1}{\pi}\,
  \mathcal{P}
  \int_{-\infty}^{\infty}
  ds\,
  \frac{\Re S(\beta+is)}{s-t}.
  \label{eq:KK-standard}
\end{equation}

However, a Fisher zero $z_k=\beta_k+i t_k$, at fixed $\beta$, corresponds to the complex-time
singularity $w_k=t_k+i(\beta-\beta_k).$ Hence, zeros with $\beta_k>\beta$ lie in the lower half-plane and the naive KK relation cannot be applied.

In this case, one separates $ F_\beta(w)
=
F_{\mathrm{reg}}(w)+F_{\mathrm{sing}}(w),$ where $F_{\mathrm{sing}}$ contains the pole and logarithmic contributions
associated with Fisher zeros. The regular part obeys the usual
Kramers--Kronig relation, while the full TPE satisfies
\begin{equation}
  \Im F_\beta(t)
  =
  \frac{1}{\pi}\,
  \mathcal{P}
  \int_{-\infty}^{\infty}
  ds\,
  \frac{\Re F_\beta(s)}{s-t}
  +
  \Delta_{\mathrm F}(t)
  ,
  \label{eq:KK-generalized}
\end{equation}
where $\Delta_{\mathrm F}(t)$ contains the information about the Fisher singularities. Thus, the real and imaginary
parts are related by a Hilbert transform only in a zero-free analytic
domain; globally, Fisher zeros supply additional phase and winding
information not contained in the naive Kramers--Kronig reconstruction.

In Figure~\ref{fig:kk-relation}, we test Eq.~\eqref{eq:KK-standard} for both the XXZ chain and the mixed-field Ising model at $N = 8$. We find that, at $\beta=0.2$, both
chains possess Fisher zeros with $\beta_k>\beta$. The naive KK reconstruction then fails, with relative errors $1.797$ for the XXZ chain and $1.668$ for the mixed-field Ising chain; the narrow divergences of $S$ at the zeros are clipped in (a) for
visibility. At $\beta=3.0$ the contour is free of the relevant zeros for both
models, and the reconstruction works. The imaginary part is then recovered from the real part alone, up to the additive constant fixed by $\Im S(\beta)=0$ at $t=0$. This transition, however, doesn't diagnose quantum chaos as it is governed entirely by the position of Fisher zeros relative to the integration line.

\section{Pseudo entropy and spectral complexity}
\label{sec:pseudo_entropy_complexity}

In this section, we will relate pseudo entropy to the spectral complexity measure, $\mathcal{C}_S(t)$, that probes quantum chaos and is also supposed to represent the holographic dual of the volume of the Einstein-Rosen bridge in the boundary theory \cite{Iliesiu:2021ari, He_2025, Balasubramanian:2019wgd, Balasubramanian:2021mxo, Erdmenger:2023wjg}. This could provide a potential holographic interpretation of thermal pseudo entropy. $\mathcal{C}_S(t)$ is defined as:

\begin{equation}
  \mathcal{C}_S(t)
  =
  \frac{2}{\tilde d\,Z(2\beta)}
  \sum_{m\neq n}
  e^{-\beta(E_m+E_n)}
  \frac{
    1-\cos\!\left((E_m-E_n)t\right)
  }{
    (E_m-E_n)^2
  } .
  \label{eq:spectral_complexity_definition}
\end{equation}
where $\tilde d$ is the Hilbert space dimension. Taking a second time derivative of $C_S(t)$ relates spectral complexity with the unnormalized spectral factor as:
\begin{align}
  \frac{d^2\mathcal{C}_S(t)}{dt^2}
  &=
  \frac{2}{\tilde d\,Z(2\beta)}
  \sum_{m\neq n}
  e^{-\beta(E_m+E_n)}
  \cos\!\left((E_m-E_n)t\right) \nonumber \\
  &=  \frac{2}{\tilde d\,Z(2\beta)}
  |Z(\beta+i t)|^2
  -
  \frac{2}{\tilde d} \nonumber \\
  &=
  \frac{2}{\tilde d\,Z(2\beta)}
  \exp\Biggl\{
    2\operatorname{Re}
    S\!\left(\mathcal{T}_{R}^{\psi|\varphi}\right)
    -
    2\operatorname{Re}
    \Bigl[
      (\beta+i t)\,
      \langle H_R\rangle_{\beta+i t}
    \Bigr]
  \Biggr\}
  -
  \frac{2}{\tilde d}.
  \label{eq:complexity_from_pseudo_entropy_final}
\end{align}
Since $\mathcal{C}_S(0)=0$ and $\left.\frac{d\mathcal{C}_S(t)}{dt}\right|_{t=0}=0$, Eq.~\eqref{eq:complexity_from_pseudo_entropy_final} can be double integrated to find the expression for spectral complexity in terms of pseudo entropy:

\begin{align}
  \mathcal{C}_S(t)
  &=
  \frac{2}{\tilde d\,Z(2\beta)}
  \int_0^t d\tau\,
  (t-\tau)\,
  \exp\Biggl\{
    2\operatorname{Re}
    S\!\left(\mathcal{T}_{R}^{\psi|\varphi}(\tau)\right)
    -
    2\operatorname{Re}
    \Bigl[
      (\beta+i\tau)
      \langle H_R\rangle_{\beta+i\tau}
    \Bigr]
  \Biggr\}
  -
  \frac{t^2}{\tilde d}.
  \label{eq:integrated_complexity_from_pseudo_entropy_final}
\end{align}
This double integration exactly gives the the $(E_m-E_n)^{-2}$ weight in
Eq.~\eqref{eq:spectral_complexity_definition}

\subsection{Spectral complexity and relative pseudo entropy}
\label{sub sec: SC and RPE}

\begin{figure*}[htbp]
  \centering
  \includegraphics[width=\textwidth]{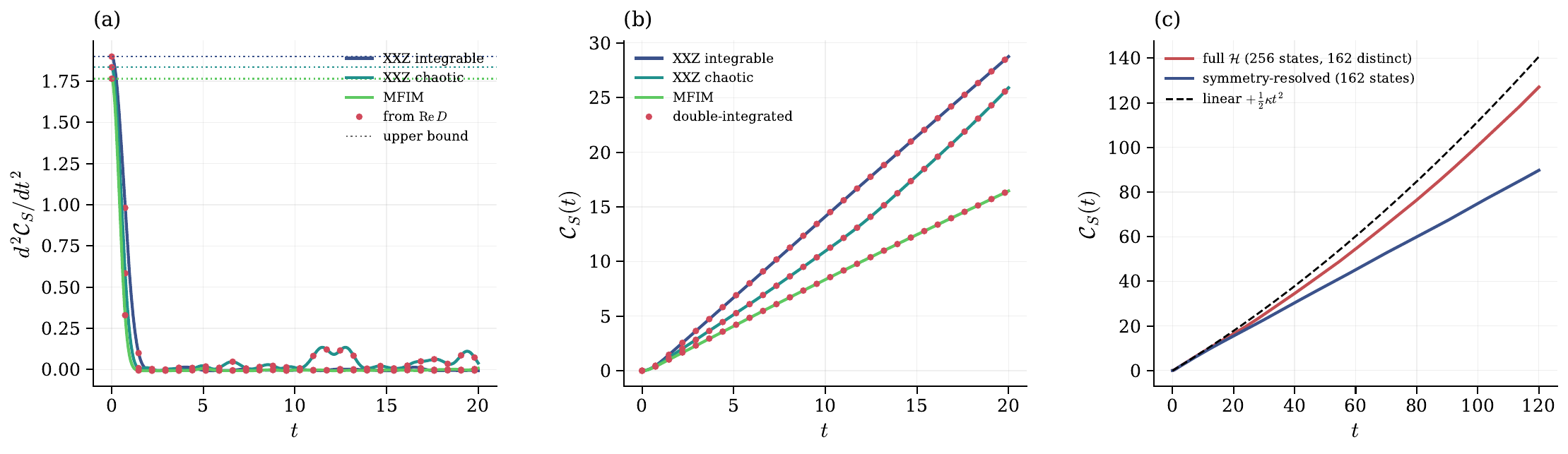}
  \caption{ For $N=8$, $\beta=0.2$. (a) Solid lines: $d^2\mathcal{C}_S/dt^2$ evaluated directly from the off-diagonal double sum in
    Eq.~\eqref{eq:spectral_complexity_definition}. Red points: the same quantity obtained from
  Eq.~\eqref{eq:complexity_from_relative_pseudo_entropy_final} using only $\operatorname{Re}D(\rho_R\|\mathcal{T}_R(t))$. Dotted lines: the upper bound of Eq.~\eqref{eq:complexity-bound}, saturated at $t=0$. (b) $\mathcal{C}_S(t)$ from the definition (solid) compared with the double-integrated form Eq.~\eqref{eq:integrated_complexity_from_relative_pseudo_entropy_final} (points). (c) Effect of spectral degeneracy for the mixed-field Ising chain: evaluated on the full Hilbert space (red) and on the degeneracy-resolved spectrum (blue), with the dashed curve showing a linear growth plus the $\tfrac12\kappa t^{2}$ term predicted by Eq.~\eqref{eq:late-time-plateau}.}
  \label{fig:complexity-check}
\end{figure*}

A more direct relation can be obtained by replacing the pseudo entropy with the relative pseudo entropy between
\begin{equation}
  \mathcal{T}_R(t)=\frac{e^{-(\beta+it)H_R}}{Z(\beta+it)},
  \qquad
  \rho_R=\frac{e^{-\beta H_R}}{Z(\beta)},
\end{equation}
We define
\begin{align}
  D\!\left(\rho_R\middle\|\mathcal{T}_R(t)\right)
  &=
  \operatorname{Tr}
  \left[
    \rho_R
    \left(
      \log\rho_R-\log\mathcal{T}_R(t)
    \right)
  \right]
  \nonumber\\
  &=
  \operatorname{Tr}\!\Bigl[
    \rho_R\bigl(
      -\beta H_R-\log Z(\beta)\,\mathbf{1}_R
      +(\beta+it)H_R+\log Z(\beta+it)\,\mathbf{1}_R
    \bigr)
  \Bigr]
  \nonumber\\
  &=
  it\,\langle H_R\rangle_\beta
  +\log Z(\beta+it)-\log Z(\beta),
  \label{eq:relative_pseudo_entropy_algebra}
\end{align}
where $\langle H_R\rangle_\beta=\operatorname{Tr}(\rho_R H_R)$ and we
used $\operatorname{Tr}\rho_R=1$. Taking the real part gives,
\begin{align}
  \operatorname{Re}D\!\left(\rho_R\middle\|\mathcal{T}_R(t)\right)
  &=
  \operatorname{Re}\log Z(\beta+it)-\log Z(\beta)
  \nonumber\\
  &=
  \log|Z(\beta+it)|-\log Z(\beta)
  \nonumber\\
  &=
  \tfrac{1}{2}\log|Z(\beta+it)|^2
  -\tfrac{1}{2}\log Z(\beta)^2 .
  \label{eq:real_relative_pseudo_entropy}
\end{align}
The real part is therefore sign-definite. Writing
$Z(\beta+it)=\sum_n e^{-\beta E_n}e^{-itE_n}$ and applying the triangle
inequality,
\begin{equation}
  \bigl|Z(\beta+it)\bigr|
  \;\le\;
  \sum_n e^{-\beta E_n}
  \;=\;
  Z(\beta),
  \label{eq:triangle}
\end{equation}
with equality precisely when the phases $e^{-itE_n}$ coincide for all
$n$. For a generic spectrum this happens only at $t=0$; for a
commensurate spectrum it recurs at the exact revival times.
Eqs.~\eqref{eq:real_relative_pseudo_entropy}
and~\eqref{eq:triangle} give
\begin{equation}
  \operatorname{Re}
  D\!\left(\rho_R\middle\|\mathcal{T}_R(t)\right)\le 0
  \;
  \label{eq:reD-nonpositive}
\end{equation}
Therefore, the real part of pseudo entropy doesn't impose the usual Klein's inequality and the data-processing inequality. In fact, it is the opposite.

Multiplying Eq.~\eqref{eq:real_relative_pseudo_entropy} by two and exponentiating,
\begin{equation}
  \exp\!\left[
    2\operatorname{Re}
    D\!\left(\rho_R\middle\|\mathcal{T}_R(t)\right)
  \right]
  =
  \frac{|Z(\beta+it)|^2}{Z(\beta)^2}\;\le\;1 .
  \label{eq:relative_pseudo_entropy_normalized_sff}
\end{equation}
Substituting into the second derivative of spectral complexity yields
the exact relative-pseudo entropy--complexity relation
\begin{equation}
  \frac{d^2\mathcal{C}_S(t)}{dt^2}
  =
  \frac{2Z(\beta)^2}{\tilde d\,Z(2\beta)}
  \exp\!\left[
    2\operatorname{Re}
    D\!\left(\rho_R\middle\|\mathcal{T}_R(t)\right)
  \right]
  -
  \frac{2}{\tilde d}.
  \label{eq:complexity_from_relative_pseudo_entropy_final}
\end{equation}
Unlike the pseudo entropy form of
Eq.~\eqref{eq:complexity_from_pseudo_entropy_final}, this expression
involves no separate $\langle H_R\rangle_{\beta+it}$ term. In this sense
$\operatorname{Re}D$ acts as the logarithmic generator of the
acceleration of spectral complexity,
\begin{equation}
  \operatorname{Re}
  D\!\left(\rho_R\middle\|\mathcal{T}_R(t)\right)
  =
  \frac{1}{2}
  \log\!\left[
    \frac{\tilde d\,Z(2\beta)}{2Z(\beta)^2}
    \left(
      \frac{d^2\mathcal{C}_S}{dt^2}+\frac{2}{\tilde d}
    \right)
  \right].
\end{equation}
An immediate consequence of Eq.~\eqref{eq:reD-nonpositive} is a purely thermodynamic bound on the growth of spectral complexity
\begin{equation}
  \;
  \frac{d^2\mathcal{C}_S(t)}{dt^2}
  \;\le\;
  \frac{2}{\tilde d}
  \left[
    \frac{Z(\beta)^2}{Z(2\beta)}-1
  \right]
  \;
  \label{eq:complexity-bound}
\end{equation}
saturated at $t=0$ and, for commensurate spectra, at the revival times.
The maximal acceleration of spectral complexity is thus fixed by
the thermal participation ratio of the spectrum. Finally, using $\mathcal{C}_S(0)=0$ and
$\left.d\mathcal{C}_S/dt\right|_{t=0}=0$,
Eq.~\eqref{eq:complexity_from_relative_pseudo_entropy_final} integrates
twice to
\begin{align}
  \mathcal{C}_S(t)
  &=
  \frac{2Z(\beta)^2}{\tilde d\,Z(2\beta)}
  \int_0^t d\tau\,(t-\tau)\,
  \exp\Biggl[
    2\operatorname{Re}
    D\!\left(\rho_R\middle\|\mathcal{T}_R(\tau)\right)
  \Biggr]
  -
  \frac{t^2}{\tilde d}.
  \label{eq:integrated_complexity_from_relative_pseudo_entropy_final}
\end{align}

There are some interesting comments left. First, both
Eqs.~\eqref{eq:complexity_from_pseudo_entropy_final}
and~\eqref{eq:complexity_from_relative_pseudo_entropy_final}  are degenerate
at the Fisher zeros, where
$Z(\beta+it)$ vanishes and
$\operatorname{Re}D\to-\infty$. Then, the complexity acceleration should
attains its minimum $-2/\tilde d$. Second, taking $t\to\infty$ in
Eq.~\eqref{eq:complexity_from_relative_pseudo_entropy_final} and
averaging over the late-time fluctuations gives
\begin{equation}
  \left\langle
  \frac{d^2\mathcal{C}_S}{dt^2}
  \right\rangle_{t\to\infty}
  =
  \frac{2}{\tilde d\,Z(2\beta)}
  \sum_j g_j^{\,2}e^{-2\beta E_j}
  -
  \frac{2}{\tilde d},
  \label{eq:late-time-plateau}
\end{equation}
where $g_j$ is the degeneracy of the distinct level $E_j$. This vanishes
if and only if the spectrum is non-degenerate. On a Hilbert space
with exact symmetry degeneracies, the complexity therefore has a
spurious contribution growing as $t^2$ without bound. This suggests that,
$\mathcal{C}_S$ must be evaluated within a single symmetry sector.

Figure~\ref{fig:complexity-check} verifies the relations of this section
for the integrable and chaotic XXZ chains and the mixed-field Ising
chain up to machine precision. We have omitted the pseudo entropy form in the figure, though numerical verification shows agreement up to the same precision. The bound \eqref{eq:complexity-bound} is saturated at $t=0$ in every
case. Panel~(c) illustrates the role of degeneracy.  Spectral
complexity is therefore meaningful only within a single symmetry sector.

\subsection{Operational meaning}

We will now try to provide an operational meaning for this relation between relative pseudo entropy and spectral complexity \eqref{eq:complexity_from_relative_pseudo_entropy_final}. The real part of the relative pseudo entropy is just the logarithm
of the survival probability of the thermofield double:
\begin{equation}
  P(t)
  \;\equiv\;
  \bigl|\braket{\mathrm{TFD}(\beta)}{\mathrm{TFD}(\beta,t)}\bigr|^{2}
  =
  \exp\!\left[
    2\operatorname{Re}
    D\!\left(\rho_R\middle\|\mathcal{T}_R(t)\right)
  \right].
  \label{eq:tfd-survival}
\end{equation}
Then, non-positivity \eqref{eq:reD-nonpositive} is the statement that a fidelity
cannot exceed unity, with equality only when the state has not yet evolved. It is also related to the Loschmidt rate function used to detect dynamical quantum phase transitions:
\begin{equation}
  \lambda(t)=-\log P(t)=-2\operatorname{Re}D.
\end{equation}

\section{Conclusion}
\label{sec:conclusion}

In this work, we investigated pseudo entropy as a probe of quantum chaos in quantum spin chains. The models considered are the Heisenberg XXZ model and the mixed-field Ising model, whose integrable and chaotic regimes can be distinguished through random-matrix level statistics. First, we establish a direct relation between thermal pseudo entropy and the spectral form factor (SFF), showing that its spectral contribution inherits the characteristic dip--ramp--plateau structure of chaotic spectra. We further clarified the imaginary part of thermal pseudo entropy: Kramers--Kronig reconstruction holds in zero-free analytic domains, while Fisher zeros introduce additional pole, phase, and winding information. Therefore, it is possible that the imaginary part carries extra information that the real part doesn't. For spatial pseudo-entropy in the XXZ chain, we found that the logarithmic critical scaling holds in both the integrable and chaotic regimes. The conjectured non-positivity bound holds throughout the critical phase but can be violated during the transition to the Néel phase. This is expected. Finally, motivated by the connections to wormholes \cite{Iliesiu:2021ari}, we derived exact relations connecting pseudo entropy, relative pseudo entropy, and spectral complexity. In particular, the real part of the relative pseudo-entropy serves as the logarithmic generator of the acceleration of spectral complexity.

Several interesting extensions are natural. For oppositely ordered states
$\ket{\psi_1(t)}=W(t)V\ket{\psi_0}$ and
$\ket{\psi_2(t)}=VW(t)\ket{\psi_0}$, their overlap is the out-of-time-order correlator, suggesting that the associated pseudo entropy may provide an entropic probe of operator scrambling, with its singularities governed by the same Fisher-zero structure discussed above. Connections with Krylov measures would also be quite interesing as they probes different time-limits \cite{Adhikari:2025vdl,Adhikari_2024}. In Quantum Field Theories (QFT), the algebraic formulation of relative entropy motivates a corresponding study of relative pseudo entropy \cite{He_2025,Adhikari:2022whf}. Holographically, it would be interesting to explore these relations for eternal black holes, complexity growth, and wormholes \cite{Cotler:2016fpe,doi:10.1142/S0217751X25480057,Adhikari_2022}. Further, JT gravity \cite{Turiaci:2024cad, TEITELBOIM198341, JACKIW1985343} is the simplest toy model of quantum gravity in $(1+1)$d, which is solvable, and its connection with RMT \cite{Stanford:2019vob} may hint at a potential connection to pseudo entropy. The framework may also be useful in quantum simulators as a diagnostic of many-body dynamics and coherent errors \cite{e27111165,adhikari2026certifiedfidelitysusceptibilityclassical,Adhikari:2026srf}.

\acknowledgments
This research is part of the Abdus Salam International Centre for Theoretical Physics (ICTP) program: Physics Without Frontiers (PWF) and we acknowledge support from the PWF program of the ICTP, Italy.
The research is part of the Munich Quantum Valley, which is supported by the Bavarian state government with funds from the Hightech Agenda Bayern Plus.
\bibliographystyle{JHEP}
\bibliography{ref4}

\clearpage
\onecolumngrid

\appendix

\end{document}